\documentclass[prl,twocolumn,amsmath,superscriptaddress,showpacs]{revtex4}
\usepackage[dvips]{graphicx,color}
\usepackage{psfrag}

\usepackage{color}
\definecolor{darkgreen}{rgb}{0,0.5,0}
\definecolor{blue}{rgb}{0,0,0.8}
\definecolor{lightblue}{rgb}{0.93,0.96,1}
\definecolor{darkblue}{rgb}{0.,0.,0.6}
\usepackage[colorlinks,linkcolor=darkgreen,citecolor=darkblue,urlcolor=blue]{hyperref}

\def\reff#1{(\ref{#1})}

\begin{document}

\title{Superexponential droplet fractalization as a hierarchical formation of dissipative compactons}
\date{\today}
\author{Sergey~Shklyaev}
\affiliation{Department of Theoretical Physics, Perm State University, 15 Bukirev St., Perm 614990,
Russia}\affiliation{Department of Physics and Astronomy, University of Potsdam, Karl-Liebknecht-Str. 24/25, D-14476
Potsdam, Germany}

\author{Arthur V. Straube}
\affiliation{Department of Physics, Humboldt University of Berlin, Newtonstr. 15, D-12489 Berlin, Germany}

\author{Arkady Pikovsky}
\affiliation{Department of Physics and Astronomy, University of Potsdam, Karl-Liebknecht-Str. 24/25, D-14476 Potsdam,
Germany}

\begin{abstract}
We study the dynamics of a thin film over a substrate heated from below in a
framework of a strongly nonlinear one-dimensional Cahn-Hillard equation. The evolution leads
to a fractalization into smaller and smaller scales. We demonstrate that
a primitive element in the appearing hierarchical structure is a dissipative compacton.
Both direct simulations and the analysis of a self-similar solution show that the compactons appear at
superexponentially decreasing scales, what means vanishing dimension of the fractal.
\end{abstract}


%05. Statistical physics, thermodynamics, and nonlinear dynamical systems:
%05.45.-a Nonlinear dynamics and chaos
%05.45.Df Fractals

%47. Fluid dynamics:
%47.15.gm Thin film flows
%47.20.Dr Surface-tension-driven instability
%47.20.Ky Nonlinearity, bifurcation, and symmetry breaking
%47.20.Ma Interfacial instabilities (e.g., Rayleigh-Taylor)
%47.53.+n Fractals in fluid dynamics
%47.54.-r Pattern selection; pattern formation
%47.55.pf Marangoni convection

%68. Surfaces and interfaces; thin films and nanosystems (structure and nonelectronic properties):
%68.03.Kn Dynamics (capillary waves)
%68.08.-p Liquid-solid interfaces
%68.08.Bc Wetting
%68.15.+e Liquid thin films

% lubrication theory, thin films, surface tension, fourth-order diffusion equations
% AMS Subject classification:
% 35-02, 35A20, 35B40, 35G20, 35K55, 35K65, 76-02, 76B45, 76D08

\date{\today}

\maketitle

{\em Introduction.}---A vast number of intriguing pattern-formation phenomena can be described with high-order nonlinear diffusion equations of Cahn-Hillard type. Since their introduction \cite{Cahn-Hilliard-58}, these equations have been successfully
applied to a great variety of natural and technological processes such as phase separation in binary mixtures,
alloys, glasses, and polymer solutions (see, e.g., surveys~\cite{{Novick-etal-and-book}}), topology transitions in a
Hele-Shaw cell \cite{Goldstein-etal}, dynamics of layered systems \cite{Lyubimov-Shklyaev-04}, thin films
\cite{thin-film-surveys}, competition and exclusion of biological groups \cite{Cohen-Murray-81}, and aggregation of aphids on leaves \cite{Lewis-94}. In the thin film context, numerical studies of an amplitude  equation of Cahn-Hillard type \cite{VanHook-etal-95-97, Oron-00}  have evidenced film rupture leading to the  formation of a cascade of ``drops'' and  ``fractal-like fingering'' \cite{Yeo-etal-03} comprising the gaps or ``dry  spots'' \cite{VanHook-etal-95-97} between the drops. These findings have been supported by direct simulations of the Navier-Stokes equations \cite{Boos-Thess-99}.

The goal of this paper is to describe this cascade as a hierarchical formation of dissipative compactons. Compacton is a
well-known compact (i.e. with finite support) traveling-wave solution, which emerges in  conservative systems with {\em nonlinear
dispersion} \cite{Rosenau}. Its stationary analogue with compact support appears  in dissipative systems with {\em nonlinear
dissipation} and, therefore, can be referred to as a stationary ``Dissipative Compacton'' (DC). Below we demonstrate that a DC is
a primitive element mediating the formation of {\em hierarchical fractal} structure, and characterize the fractal properties of
this structure quantitatively.

We focus on a one-dimensional Cahn-Hilliard equation describing dissipative evolution of a conserved field $h(x,t)$
\begin{equation}\label{CHEq}
h_t + \left[f(h) h_x + g(h) h_{xxx}\right]_x=0\;.
\end{equation}
\noindent In the context of the dynamics of a thin film over a substrate heated from below, this equation describes a
surface-tension-driven convection (see, e.g., Eq.~(4) in Ref.~\cite{Oron-00}), where $h$ is the local thickness and
\begin{equation} \label{funs_fg}
f=-{\rm Bo}\, h^3+ \frac{BMh^2}{2\left(1+Bh\right)^2}, \quad g=h^3.
\end{equation}
Dimensionless Bond (${\rm Bo}$), Biot ($B$), and Marangoni ($M$)
numbers determine the levels of the gravity, of the heat flux at
the free surface, and of the convective flow, respectively.
Although function $f$ here has a rather complex form, for $h\to 0$
one can set $f\approx 0.5BM\, h^2$. This approximation holds also
for moderate values of $h$, provided the gravity can be neglected,
${\rm Bo}\to 0$, the heat transfer at the free surface is poor
($B$ small) while the thermocapillary effect is strong ($M$
large). The function $g(h)$, which can be referred to as
``mobility'', is conventionally non-negative, $g(h)\ge 0$, as this
prevents against the fast growth of the short-wave perturbations.

Assuming the limiting form of $f$, after an appropriate rescaling of the time, the field,
and the coordinate we  arrive at our basic equation
\begin{equation}\label{simple-CH}
h_t+\left(h^2h_x +  h^3h_{xxx}\right)_x=0.
\end{equation}
\noindent Noteworthy, Eq.~\reff{simple-CH} is invariant under the scaling
\begin{equation}\label{transf}
h \to p^2 h, \quad x \to px, \quad t \to p^{-2} t\;,
\end{equation}
\noindent meaning that a thinner film evolves slower.

\begin{figure}[!tb]
\includegraphics[width=0.46\textwidth]{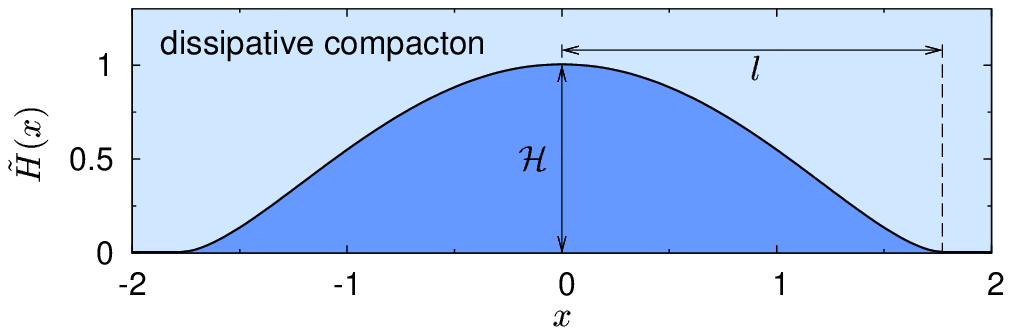}
\caption{(Color online). The shape of the base DC, $\tilde H(x)$.} \label{fig:DC-profile}
\end{figure}

{\em Steady state and its stability.}---We start our analysis by considering positive stationary solutions $h=H(x)$ of Eq.~\reff{simple-CH}. Looking for symmetric patterns, after one integration we obtain
\begin{equation}
H H^{\prime\prime\prime}+H^\prime=0 \quad \label{ode}
\end{equation}
\noindent with primes denoting $d/dx$. Equation \reff{ode} admits a compact solution $H(x)$ in the form of a DC or a
``touchdown steady state'' \cite{Laugesen-Pugh-00}, nonvanishing for $|x| \le l$ only:
\begin{equation}
x=\pm \sqrt{\pi \mathcal{H}}\,{\rm erf}\left(\sqrt{\frac{1}{2}\ln \frac{\mathcal{H}}{H}}\right),  \quad
\mathcal{H}=\max_x H(x), \label{x(h)}
\end{equation}
\noindent where ${\rm erf}(z)=\sqrt{2/\pi}\int_0^z e^{-t^2}{\rm d}t$. Solution \reff{x(h)} presents a self-affine
one-parameter family of DCs parametrized by $\mathcal{H}$. For a thin film, the DC describes the stationary profile of a drop with the amplitude $\mathcal{H}$ and zero contact angle. Owing to scaling \reff{transf}, any DC can be expressed in terms of the base DC $\tilde{H}(x)$ having $\mathcal{H}=1$, see Fig.~\ref{fig:DC-profile}. Thus, the profile of a DC and half its length $l$ obey:
\begin{equation}
H(x)=\mathcal{H} \,\tilde{H}\left(x/\sqrt{\mathcal{H}}\right), \quad l=\sqrt{\pi\mathcal{H}}\;. \label{DCfamily} \\
\end{equation}

The property of self-affinity is a necessary prerequisite for the
emergence of fractal structure described by Eq.~\reff{simple-CH},
as we discuss below. As it follows from Eq.~\reff{DCfamily}, DCs
become narrower for smaller amplitudes -- contrary to other examples
of compactons where typically the width is
amplitude independent~\cite{Rosenau}. In a more general situation,
when for small $h$ the functions $f(h)$ and $g(h)$ scale like
$f/g\propto h^{\gamma}$, no solutions are possible for $\gamma \le
-2$ \cite{Laugesen-Pugh-00}. At $\gamma > 0$, the solutions are
soliton-like because their support is no more compact. The case
$\gamma=0$ results in constant $l$. Compact solutions satisfying
the requirement of $l \to 0$ as $\mathcal{H}\to 0$ exist in the
range of $-2 < \gamma < 0$ only. Thus, the considered above case
$\gamma=-1$ is the only integer index possessing self-affine
compactons.

%%%%%%%%%%%%%%%%%%%%%%%%%%%%%%%%%%%%%%%%%%%%%%%%%%%%%%%%%%%%%%%%%%%%%%%%%%%%%%%%%
%
%
%%%%%%%%%%%%%%%%%%%%%%%%%%%%%%%%%%%%%%%%%%%%%%%%%%%%%%%%%%%%%%%%%%%%%%%%%%%%%%%%%

To explore the stability of a DC, we introduce a small perturbation $\propto \xi(x)\exp(\lambda t)$ of $H(x)$, where $\lambda$ is the growth rate. By linearizing Eq.~\reff{simple-CH}, we obtain
\begin{equation}\label{linear-stab-eq}
\lambda \xi+\left[H^3\left(\xi^{\prime\prime}+
H^{-1}\xi\right)^{\prime}\right]^{\prime}=0.
\end{equation}
\noindent Assuming $\xi(\pm l)=0$, we multiply
Eq.~\reff{linear-stab-eq} by $\xi^{\prime\prime}+ H^{-1}\xi$ and
integrate by parts to arrive at an integral relation
\begin{equation}\label{int-relation}
\lambda \int_{-l}^l
\left(\xi^\prime-\frac{H^{\prime\prime}}{H^\prime}\xi\right)^2
{\rm d} x= -\int_{-l}^l H^3
\left[\left(\xi^{\prime\prime}+\frac{\xi}{H}\right)^{\prime}\right]^2
{\rm d} x,
\end{equation}
\noindent which is closely related to the variational principle
for Eq.~\reff{CHEq} \cite{Laugesen-Pugh-02} and the fact that the
Lyapunov functional has a local minimum on the DC. As $H \ge 0$,
both the integrals in Eq.~\reff{int-relation} are non-negative and
the perturbations are nongrowing, $\lambda \le 0$. This result,
however, does not guarantee against the instability, as there
exist two modes of neutral stability, $\lambda=0$, satisfying
$\xi(\pm l)=0$. One mode, $\xi_1^{(0)}= H^\prime$, reflects
translation invariance and cannot give rise to instability.
Another mode, $\xi_2^{(0)}= H-x H^{\prime}/2$, has a nonzero
volume and can be destabilizing, if nonlinear corrections are
retained.

Thus, although a DC is stable with respect to perturbations of zero volume, the instability is possible if the volume
of the DC is changed, as confirmed by our numerical simulations of Eq.~\reff{simple-CH}. We detect the temporal decay
even for finite-amplitude perturbations of zero volume. For the perturbations changing the volume but keeping
constant the length of the DC, we find a breakup of the DC with the emergence of a complex structure.

\begin{figure}[!b]
\includegraphics[width=0.48\textwidth]{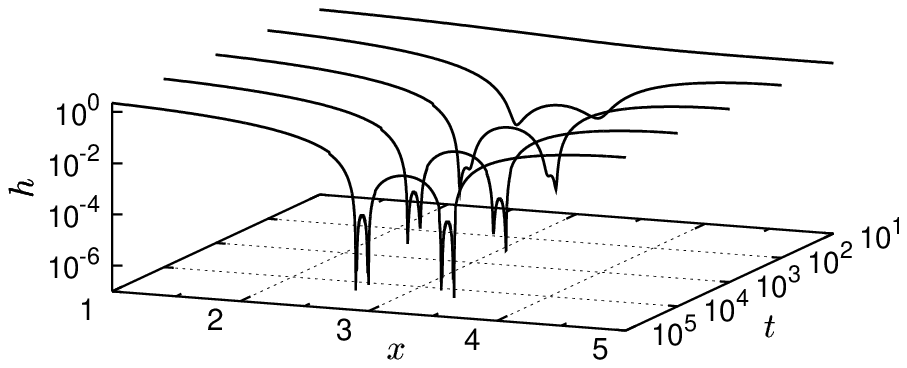}
\caption{Fragment of the evolution of the field illustrating hierarchical
formation of droplets. Notice logarithmic scales of the time and the field.}
\label{fig:evol0}
\end{figure}

\begin{figure}[!tbh]
\includegraphics[width=0.48\textwidth]{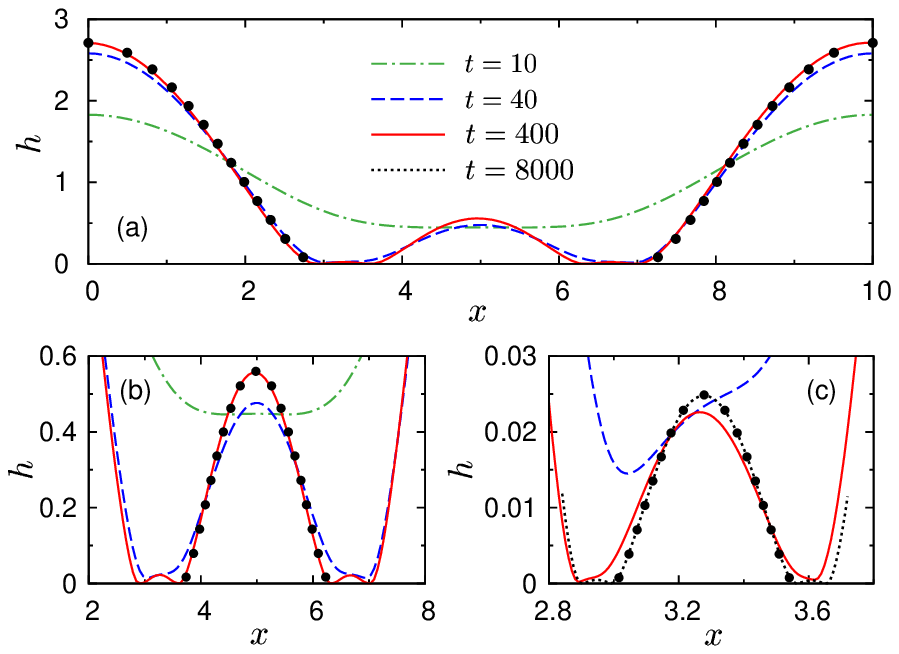}
\caption{(Color online). Evolution of the film profile for $d=10$. Panels (b) and
(c) are zoomed in fragments of panel (a). Lines represent numerical results according to Eq.~\reff{simple-CH}, circles
show the profiles of corresponding DCs as in Eq.~\reff{x(h)}. } \label{fig:evol}
\end{figure}

{\em Evolutionary problem.}---We now study the formation of a fractal, hierarchical structure of DCs, illustrated in
Fig.~\ref{fig:evol0}. We discretize Eq.~\reff{simple-CH} in the
computation domain $0 \le x \le d$ with periodic boundary conditions, with the number of nodes
$N=1000$ and apply the Newton-Kantorovich method \cite{Oron-00}.
We choose a distorted uniform profile $h(x,t=0)=1+a\cos
(2\pi x/d)$ with $a=0.1$ as an initial condition. Results of computations are presented in
Figs.~\ref{fig:evol0} and \ref{fig:evol}.  There we also compare the numerically obtained
profile $h(x)$ having local maxima $h_m^{(n)}$, $n=1, 2, \dots$ with the DC profiles with $\mathcal{H}=h_m^{(n)}$, $\text{DC}^{(n)}$ below, what indicates that the initial state develops into a hierarchical structure of DCs of different amplitudes.
%
%%%%%%%%%%%%%%%%%%%%%%%%%%%%%%%%%%%%%%%%%%%%%%%%%%%%%%%%%%%%%%%%%%%%%%%%%%%%%%%%%
%%%%%%%%%%%%%%%%%%%%%%%%%%%%%%%%%%%%%%%%%%%%%%%%%%%%%%%%%%%%%%%%%%%%%%%%%%%%%%%%%

This fact allows us to increase the efficiency of the numerics significantly: because after their formation
the DCs remain constant in their bulk, we exclude these domains from numerical simulations and impose the corresponding boundary
conditions for the still evolving domains between the formed DCs. Thus, while proceeding to smaller DCs,
we can considerably refine the mesh and also increase the time step. Therefore, we not only resolve high-order DCs with the accuracy
consistent with that at previous stages, but also maintain the computational efficiency. This strategy provides reliable results up to $n=4$.

%%%%%%%%%%%%%%%%%%%%%%%%%%%%%%%%%%%%%%%%%%%%%%%%%%%%%%%%%%%%%%%%%%%%%%%%%%%%%%%%%
%%%%%%%%%%%%%%%%%%%%%%%%%%%%%%%%%%%%%%%%%%%%%%%%%%%%%%%%%%%%%%%%%%%%%%%%%%%%%%%%%

\begin{figure}[!b]
\includegraphics[width=0.48\textwidth]{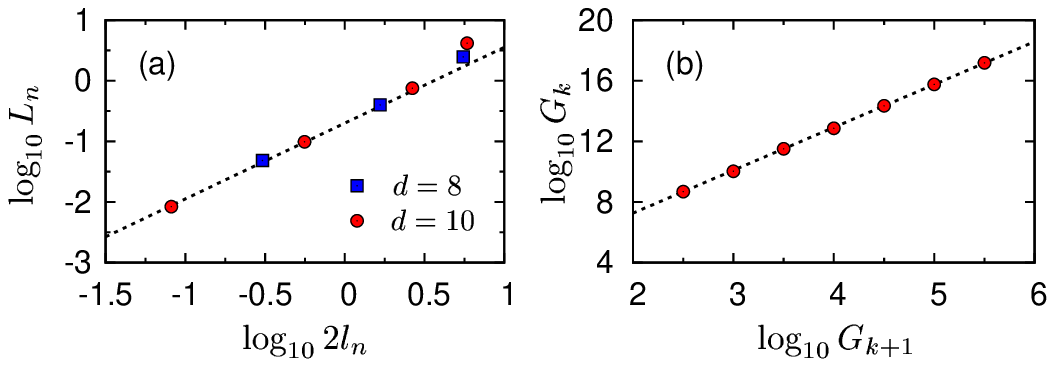}
\caption{(Color online). (a) The distance $L_n$ between two neighboring $\text{DC}^{(n)}$ and $\text{DC}^{(n-1)}$ versus the base $2l_n$
of $\text{DC}^{(n)}$. Squares and circles are numerical results for $d=8$ and $d=10$. Dotted line is a fit, Eq.~\reff{Ln_ln}. (b) Mapping $\log G_{k+1}(\log G_k)$ calculated in the framework of Eq.~\reff{matching} (circles); dotted line
corresponds to the asymptotic law, Eq.~\reff{superexp-ampl}.
} \label{fig:Ln_ln}
\end{figure}

The observed structure along with the property of self-affinity
suggests that the formation of higher-order DCs never stops and
the dry spots between DCs, form a fractal reminiscent of
the Cantor set. Here, a DC plays a role of a primitive element,
mediating the fractalization. To characterize properties of this
fractal quantitatively, we plot in Fig.~\ref{fig:Ln_ln}(a) the
variation of $L_n$, the distance between the neighboring DCs of
$n$-th and $(n-1)$-th orders, versus the base $2l_n$ of
$\text{DC}^{(n)}$. The numerical results for different $d$ fit
well a power law:
\begin{equation}\label{Ln_ln}
L_n\approx \alpha \left(2l_n\right)^{\beta}, \quad \alpha\approx 0.2, \quad \beta\approx1.25.
\end{equation}
\noindent Note that deviations from this law for the points
related to the biggest DCs stem from the initial condition.
On the other hand, for higher orders the self-similarity of the
formation of DCs is evident from Fig.~\ref{fig:Ln_ln}(a).

%%%%%%%%%%%%%%%%%%%%%%%%%%%%%%%%%%%%%%%%%%%%%%%%%%%%%%%%%%%%%%%%%%%%%%%%%%%%%%%%%
%%%%%%%%%%%%%%%%%%%%%%%%%%%%%%%%%%%%%%%%%%%%%%%%%%%%%%%%%%%%%%%%%%%%%%%%%%%%%%%%%

Because $\beta>1$ in Eq.~\reff{Ln_ln}, with the increase in $n$ the ratio $L_n/l_n$ diminishes implying that the
smaller daughter DCs tend to occupy the whole space between their bigger parent DCs. The fraction of dry spots tends
to zero and, therefore, the fractal dimension of this set equals zero. Furthermore, for large $n$ we can neglect the
distance between $\text{DC}^{(n)}$ and $\text{DC}^{(n+1)}$ and put $L_n\approx 2l_{n+1}$. As a result,
Eq.~\reff{Ln_ln} entails a remarkable superexponential scaling of $l_n$ with $n$:
\begin{equation}
\log(l_n)\propto \beta^n \log(l_0).
\label{superexp-lengths}
\end{equation}

\begin{figure}
\includegraphics[width=0.45\textwidth]{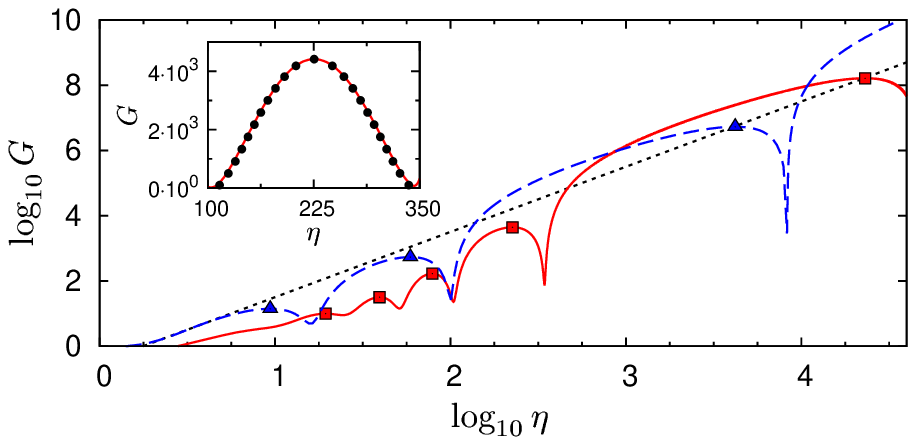}
\caption{(Color online). Function $G(\eta)$ in solution \reff{selfsim}. Initial conditions at small $\eta$ are
$G\approx 0.5\eta^2$ (solid line) and $G\approx 1-0.1 \eta^2$ (dashed line). Squares and triangles show the local maxima $G_{k}$ of $G(\eta)$; for large $k$, the maxima $G_{k}$ approach the law $G=\eta^2/\pi$ (dotted line). Inset: a comparison of a piece of $G(\eta)$ with a single DC (circles), Eq.~\reff{x(h)}.}
\label{fig:selfsim}
\end{figure}

{\em Self-similar solution.}---To shed light on the hierarchical formation of DCs and to alternatively support the conclusions about the fractal dimension and the superexponential scaling, we construct self-similar solutions, which originate from the rescaling
property, Eq.~\reff{transf}. By seeking the solution of Eq.~\reff{simple-CH} in the form
\begin{equation}\label{selfsim}
h=t^{-1}\,G(\eta), \quad \eta=x\sqrt{t},
\end{equation}
\noindent we arrive at an ordinary differential equation for $G(\eta)$:
\begin{equation}\label{F_xi}
\eta\,G'-2G+2\left(G^2G^\prime+G^3G^{\prime\prime\prime}\right)^\prime=0,
\end{equation}
\noindent where primes stand for $d/d\eta$. Numerical solutions of Eq.~\reff{F_xi} with various initial
conditions all demonstrate a qualitatively similar behavior of $G(\eta)$, which displays an infinite number of
oscillations of increasing amplitude. Two numerical solutions corresponding to different initial conditions are shown
in Fig.~\ref{fig:selfsim}.

For large $G$, where we can estimate $d/d\eta \sim \epsilon^{1/2}$ with $\epsilon\sim G^{-1} \ll 1$, the first two terms in
Eq.~\reff{F_xi} become negligible in comparison with the last two terms. In this limit, Eq.~\reff{F_xi} is reduced to Eq.~\reff{ode}
with $G(\eta)$ instead of $H(x)$. Therefore, $G(\eta)$ can be approximated by the solution for a DC [see Eq.~\reff{x(h)} and the
inset in Fig.~\ref{fig:selfsim}] everywhere except for its tails, where $G$ is no longer large. As a result, $G(\eta)$ looks like a
sequence of  DCs with superexponentially growing amplitudes $G_k\sim \exp[A^k]$ and widths $\Delta\eta_k = 2\sqrt{\pi G_k}$; the positions $\eta_k$ of maxima for large $k$ satisfy $\eta_k\approx \sqrt{\pi G_k}$, see markers in Fig.~\ref{fig:selfsim}.

To specify the superexponential growth of $G_k$ with $k$, we construct a mapping $G_k\to G_{k+1}$ valid for large
$G$. In the range of $|\eta-\eta_{k}|<\sqrt{\pi G_{k}}$, $G \approx G_{k} \tilde H(x_{k})$ with $x_k\equiv
\left(\eta-\eta_k\right)/\sqrt{G_k}$. To bridge the solution for $\text{DC}^{(k)}$ with that for $\text{DC}^{(k+1)}$,
we substitute a representation $G=\varepsilon^{-2}\zeta(y)$, $y=(\eta-\eta_k-\sqrt{\pi G_k})\varepsilon$,
$\varepsilon\equiv G_k^{-1/6}$ into Eq.~\reff{F_xi} and neglect the terms $\propto\varepsilon$,
what yields
\begin{equation}\label{matching}
y_0\,\zeta^\prime+2\left(\zeta^2\zeta^\prime+\zeta^3\zeta^{\prime\prime\prime}\right)^\prime=0.
\end{equation}
\noindent Here, primes denote $d/dy$ and
$y_0=\eta_k/\sqrt{G_k}+\sqrt{\pi}\approx 2\sqrt{\pi}$. As
Eq.~\reff{matching} admits no analytical solution, we solved it
numerically, with the initial condition at $y=-\varepsilon^{-2}(\sqrt{\pi}-x)$:
$d^p\zeta/dy^p=\varepsilon^{2(p-2)}d^p\tilde H/dx^p$  with $p=0, 1, 2, 3$ and $d^p\tilde H/dx^p$ evaluated via Eq.~\reff{x(h)} at $\mathcal H=1$, which ensures the matching with the decaying tail of $\text{DC}^{(k)}$. To perform the matching with the growing tail of $\text{DC}^{(k+1)}$ at $y\gg 1$, we take into account that $\tilde H^{\prime\prime}=-\ln \tilde H -1$ [cf. Eq.~\reff{ode}] and obtain
\begin{equation}
G_{k+1}=G_k^{1/3}\zeta\exp\left(\zeta^{\prime\prime}+1\right). \nonumber
\end{equation}
\noindent By determining $\zeta$, we end up with the transformation of $G_k\to G_{k+1}$ [see Fig.~\ref{fig:Ln_ln}(b)],
which fits well a power law
\begin{equation}\label{superexp-ampl}
G_{k+1}\approx 40\,G_k^{2.83}.
\end{equation}

Equation~\reff{superexp-ampl} shows a superexponential growth for $G_k$ with $k$, which is expected as required by the
self-affinity and the similar behavior for the lengths, see Eq.~\reff{superexp-lengths}. As the amplitude $\propto l^2$ [cf. Eq.~\reff{DCfamily}], the exponent $2.83$ in Eq.~\reff{superexp-ampl} is in reasonable agreement with
$2\beta$ in Eq.~\reff{Ln_ln}, obtained within the evolutionary problem. The fact that the correspondence is not perfect is not
surprising as Eq.~\reff{superexp-ampl} is the asymptote of extremely large $t$ (i.e. large $k$),
 while Eq.~\reff{Ln_ln} is a fit obtained for the early stage of the  evolution (small $k$).
Nevertheless, we see that the self-similar solution is closely related to the hierarchical structure of DCs described by
the evolutionary problem.

%%%%%%%%%%%%%%%%%%%%%%%%%%%%%%%%%%%%%%%%%%%%%%%%%%%%%%%%%%%%%%%%%%%%%%%%%%%%%%%%%
%%%%%%%%%%%%%%%%%%%%%%%%%%%%%%%%%%%%%%%%%%%%%%%%%%%%%%%%%%%%%%%%%%%%%%%%%%%%%%%%%

Finally, we stress that the relation between the self-similar solution
and the spatially periodic solution as in Fig.~\ref{fig:evol} is
not simple. The whole structure of successive DC-like solutions,
$h(x,t)$, obtained via Eq.~\reff{F_xi}, moves with the time toward
the point $x=0$, whereas DCs $H(x)$, which are born as a result of
evolution according to Eq.~\reff{simple-CH}, are stationary objects. However, the long-time evolutions of both these solutions
show the similar displacement of the gaps between DCs by
higher-order DCs. This argument becomes transparent, if we observe
the self-similar solution ``stroboscopically''. Let us consider a
self-similar solution at moments of time  $t_k=\eta_k^2/x_0^2$.
The corresponding field profile \reff{selfsim} describes the
formation of DCs up to the $k$-th order in the domain $0\leq
x< x_0+\Delta \eta_k/\sqrt{t_k}$ with $\text{DC}^{(k)}$ centered
at $x=x_0$. As the growth of $G_k$ with $k$ is superexponential, the highest-order DC dominates the pattern, what ensures that
the fractal made of the dry spots has zero dimension.

{\em Conclusions.}---We have applied the concept of dissipative compactons to
the evolution of a thin film within a framework of the generalized
one-dimensional Cahn-Hilliard equation. We have shown that as a result of a rupture,
the thin film evolves into a hierarchical structure of drops, which can be represented by dissipative compactons of
different scales. By efficiently solving the amplitude equation and, alternatively, by constructing a self-similar
solution, we show that this structure of DCs is a fractal, characterized by superexponentially decreasing amplitudes
and lengths of smaller droplets, and thus having zero dimension. The dissipative compacton is a primitive element
mediating the fractal structure comprising the dry spots between the compactons. It should be also noted, that a
number of effects, such as intermolecular interaction between liquid and
solid, contact angle dynamics, evaporation, etc. become of crucial
importance, when the free fluid surface touches the solid. An extension of the theory above that includes these
effects remains a challenge.

{\em Acknowledgements.}---We are grateful to A.~Nepomnyashchy, A.~Oron, M.~Zaks, Ph. Rosenau, D.~Lyubimov, and
D.~Goldobin for stimulating discussions. The research was supported by German Science Foundation (projects No.~436
RUS113/977/0-1 and No.~STR~1021/1-2) and Russian Foundation for Basic Research (project No. 08-01-91959).

\end{document}